\documentclass[12pt,showpacs,preprintnumbers,amsmath,amssymb]{revtex4}
\usepackage{graphicx}
\begin{document}

\newcommand{\pderiv}[2]{\frac{\partial #1}{\partial #2}}
\newcommand{\deriv}[2]{\frac{d #1}{d #2}}
\newcommand{\eq}[1]{Eq.~(\ref{#1})}  

\title{Ising Spin Glasses on Wheatstone-Bridge Hierarchical Lattices}

\vskip \baselineskip

\author{Octavio R. Salmon$^{1,2}$}
\thanks{E-mail address: octavior@cbpf.br}

\author{Br\'aulio T. Agostini$^{1}$}
\thanks{E-mail address: agostini@cbpf.br}

\author{Fernando D. Nobre$^{1,2}$}
\thanks{Corresponding author: E-mail address: fdnobre@cbpf.br}
\thanks{FAX: 55-21-21417401} 
\thanks{ \\ Telephone: 55-21-21417513}

\address{
$^{1}$Centro Brasileiro de Pesquisas F\'{\i}sicas \\
Rua Xavier Sigaud 150 \\
22290-180 \hspace{5mm} Rio de Janeiro - RJ \hspace{5mm} Brazil \\
$^{2}$National Institute of Science and Technology for Complex Systems \\ 
Rua Xavier Sigaud 150 \\
22290-180 \hspace{5mm} Rio de Janeiro - RJ \hspace{5mm} Brazil}

\date{\today}

\vspace{1cm}

\begin{abstract}

\begin{center}

{\large\bf Abstract}

\end{center}

\noindent
Nearest-neighbor-interaction Ising spin glasses are studied on 
three different hierarchical lattices, all of them belonging 
to the Wheatstone-Bridge family.    
It is shown that the spin-glass lower
critical dimension in these lattices should be greater than $2.32$~. 
Finite-temperature spin-glass phases are found for a lattice of fractal
dimension $D \approx 3.58$ (whose unit cell is obtained from 
a simple construction of a part of the cubic lattice), 
as well as for a lattice of fractal dimension close to five.   
In the former case, the estimates of spin-glass critical temperatures 
associated with symmetric  
Gaussian and bimodal distributions are very close to recent results
from extensive numerical simulations carried on a cubic lattice, 
suggesting that whole phase diagrams presented, obtained  for
couplings following non-centered distributions -- not known up to the
moment for Bravais lattices -- should represent good approximations.

\vspace{1cm}


\end{abstract}
\maketitle

\newpage

The Ising spin glass (SG)
\cite{dotsenkobook,nishimoribook} represents a very
controversial model in the literature of magnetic systems and has called
the attention of many workers along the last decades. Theoretically, its
simple formulation in terms of binary variables has led to a large amount
of studies from both computational and analytical points of view. The
analytical approaches have been mostly applied to its mean-field
formulation, defined in terms of infinite-range interactions. According to
its mean-field solution, the   
SG phase is properly described by an
infinite number of order parameters (i.e., an order-parameter function
\cite{parisirsb}), manifesting the property of replica-symmetry  
breaking~\cite{parisireview}.  
A major question in the Ising SG problem nowadays concerns the
identification of which properties from this mean-field solution persist in
the limit of short-range (e.g., nearest-neighbor) interactions.  
In this later case, the majority of works were concentrated
on three-dimensional Ising SG models, for which it is generally accepted
nowadays that a SG phase occurs at finite temperatures
\cite{parisireview,southernyoung,bhattyoung85,ogielski85,braymoore87,%
bhattyoung88,young04,young06,jorg,jorgkatzgraber1}; it should be 
mentioned that some of
these studies have required extensive computational efforts.   

\begin{figure}[t]
\begin{center}
\includegraphics[width=0.25\textwidth,angle=0]{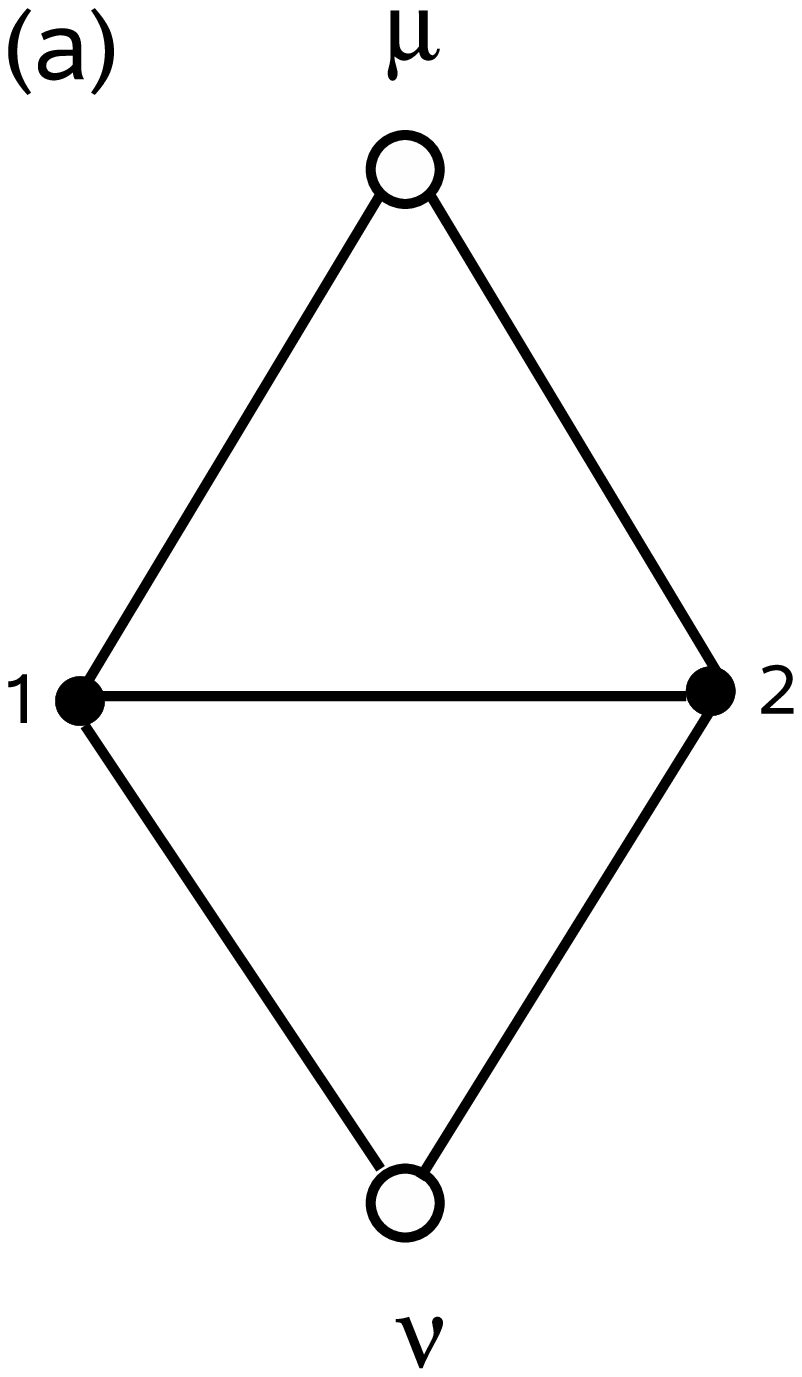}
\hspace{0.5cm}
\includegraphics[width=0.30\textwidth,angle=0]{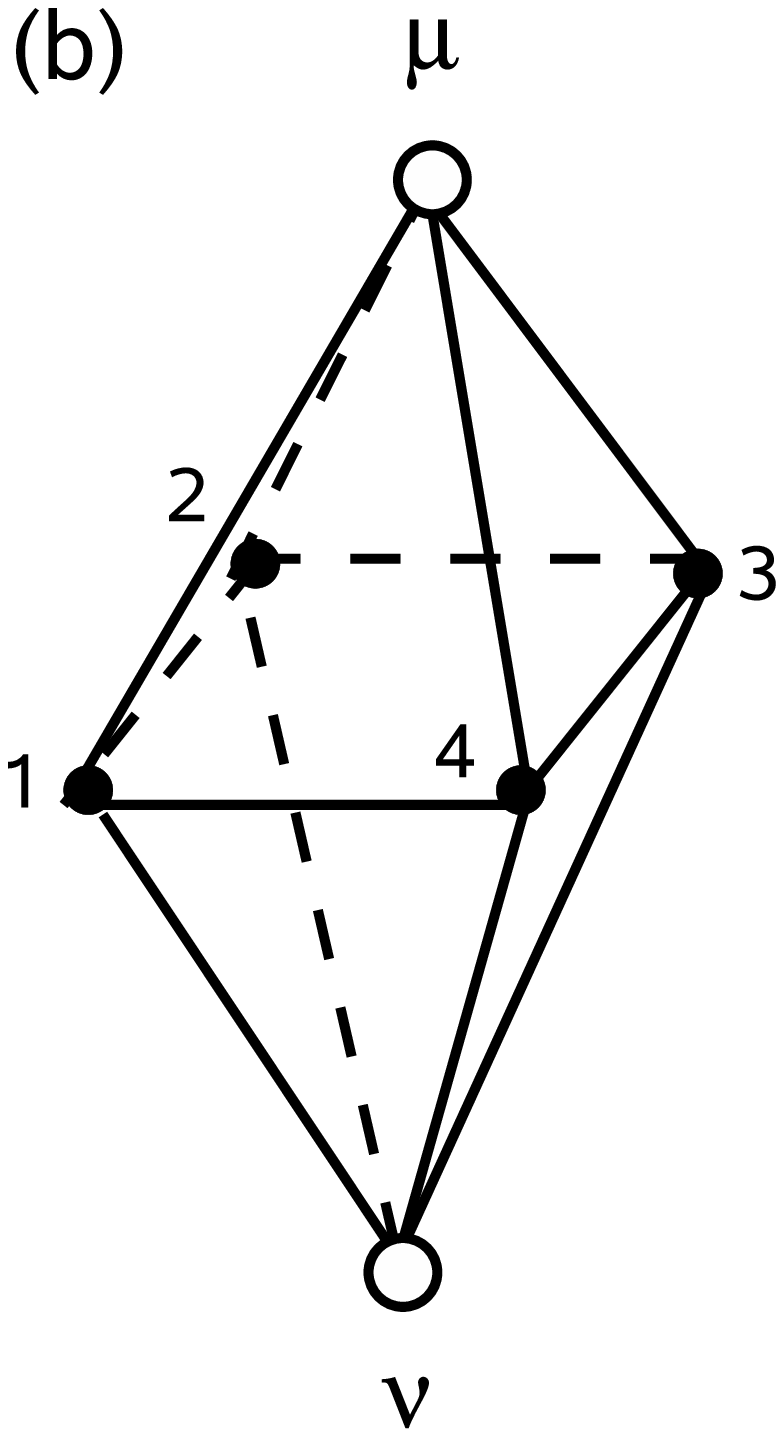}
\hspace{0.5cm}
\includegraphics[width=0.32\textwidth,angle=0]{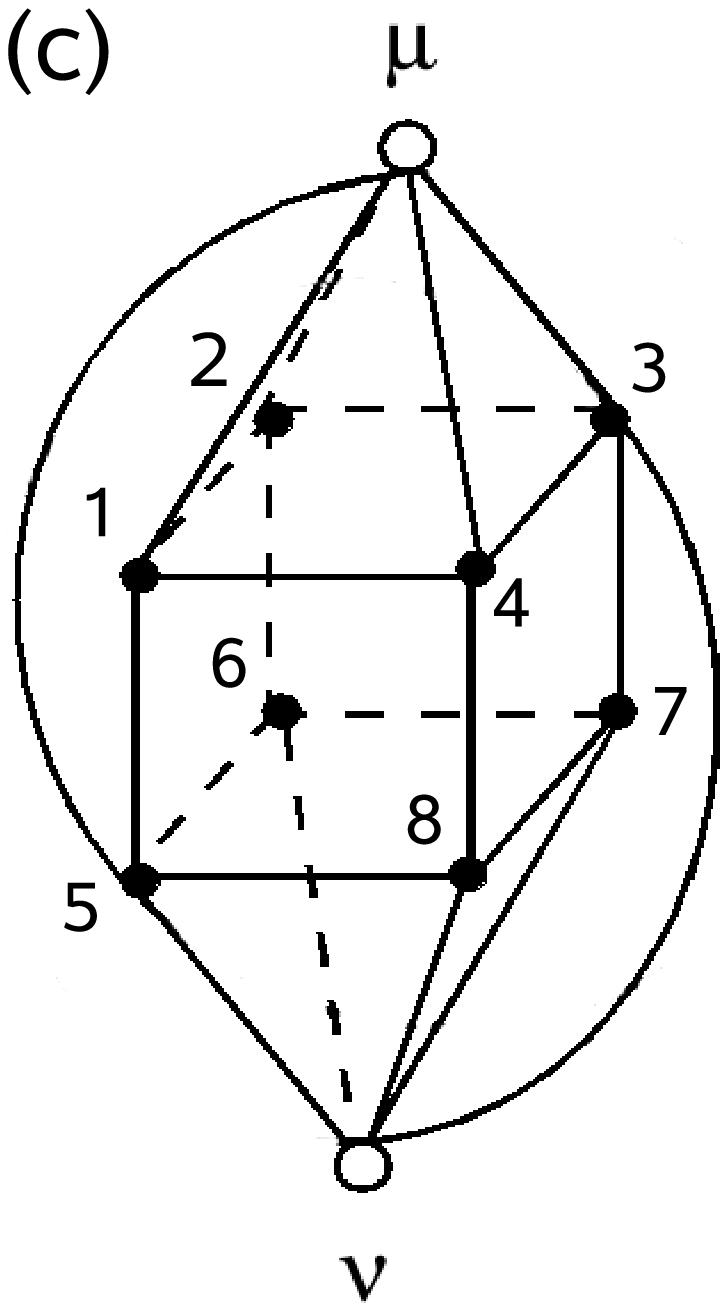}
\end{center}
\protect\caption{Basic cells of the WB family of
hierarchical lattices, with a scaling factor $b=2$.  
In each case, the empty circles ($\mu$ and $\nu$) represent the external
sites  
of the cell, whereas the black circles are internal sites to be 
decimated in the RG procedure. 
(a) The WB cell of fractal dimension $D=(\ln5)/(\ln2) \approx 2.32$; 
(b) The WB cell of fractal dimension $D=(\ln12)/(\ln2) \approx 3.58$; 
(c) The WB cell of fractal dimension $D=(\ln28)/(\ln2) \approx 4.81$. For
clearness, all the bonds of cell (c) are not shown explicitly; the external  
site $\mu$ is connected directly to sites $5,6,7,$ and $8$ (as illustrated
by the bond connecting sites $\mu$ and $5$, whereas site $\nu$ is connected
directly to sites $1,2,3,$ and $4$ (as illustrated by the bond connecting
sites $\nu$ and $3$).}
\label{fig:hlattice}
\end{figure}

\begin{figure}[t]
\begin{center}
\includegraphics[width=0.70\textwidth,angle=0]{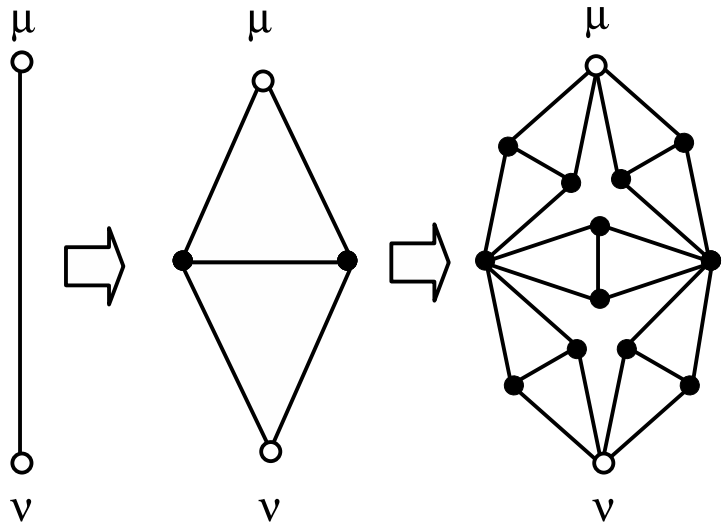}
\end{center}
\protect\caption{First three levels (levels $0,1,$ and $2$) in the
generation process of the Wheatstone-Bridge hierarchical lattice with
fractal dimension $D \approx 2.32$. The lattice-generation process starts at 
the $0$-th level with a single bond and at each step a bond is replaced by
its respective basic cell. In each level, the empty circles ($\mu$ and $\nu$) 
represent the external sites of the lattice, 
whereas the black circles are internal sites to be decimated in successive 
RG operations.}
\label{fig:latticegen2d}
\end{figure}

Hierarchical lattices 
were introduced within the context of the real-space renormalization-group 
(RG) approach, bringing the great advantage that such technique becomes 
exact for pure systems defined on these lattices~\cite{tsallismagalhaes}.
These lattices are constructed by carrying successive similar operations at
each hierarchical level, e.g., at each level one replaces bonds by
well-defined unit cells; typical examples of unit cells are presented in 
Fig.~\ref{fig:hlattice}, all of them belonging to the 
Wheatstone-Bridge~(WB) family of hierarchical lattices. Two important
characteristics of a given hierarchical lattice are its fractal dimension
$D$ and its scaling factor $b$, which is defined as the smallest distance
(counted in bonds) between the two external sites of its basic cell. The WB
hierarchical lattices are generated by starting the process 
from the $0$th level of 
the lattice-generation hierarchy
with a single bond joining the external sites (denoted by $\mu$ and $\nu$); 
then, in each iteration step one replaces a bond by its corresponding 
unit cell, in such a way that in its first 
hierarchy the lattice is represented by this unit cell. 
This procedure is shown explicitly in Fig.~\ref{fig:latticegen2d}, 
for the first three levels of the WB hierarchical lattice 
with fractal dimension $D=(\ln5)/(\ln2) \approx 2.32$. 
The process is
continued with the hierarchical lattice being constructed
up to a given ${\cal N}$-th hierarchy (${\cal N} \gg 1$).  

Although the RG may not be considered in general as an exact procedure for
random systems on hierarchical lattices, it is expected to represent a good
approximation, since in many cases, pure systems appear as particular
limits of some random models. In the particular case of Ising SGs, the
hierarchical lattices have  
been a very useful tool
\cite{southernyoung,braymoore87,banavar87,banavar88,curadomeunier,%
niflehilhorst92,nobre98,drossel98,nogueira99,curado99,%
nobre00,drossel00,nobre01,nobre03,nishimoriberker,
salmonpre09,jorgkatzgraber1},
essentially due to the possibility of performing relatively
low-time-consuming numerical computations. 
A significant gain in the computational time in the study of SGs 
represents the main advantage of
the hierarchical lattices with respect to Monte Carlo simulations performed
on Bravais lattices. In fact, one may obtain whole phase diagrams for
non-symmetric distributions of the couplings (e.g., by changing the
probability $p$ of a bimodal distribution) within typically, a few
computation hours of a standard desk computer; on the other hand, 
the criticality of a given 
particular case (e.g., the study of the case $p=1/2$ of the bimodal 
distribution for the couplings) through Monte Carlo simulations 
may require several days on high-performance computer networks.  

The Migdal-Kadanoff (MK) family of hierarchical 
lattices~\cite{migdal,kadanoff} have been widely explored,
essentially due to the fact that its basic cell is formed  
by one-dimensional parallel paths each with a given scaling factor $b$; its 
fractal dimension may be varied either by changing the scaling factor, 
or by a simple operation of adding 
more paths to the cell. As a consequence, most of the hierarchical-lattice
SG studies so far were concentrated on these lattices 
\cite{southernyoung,braymoore87,banavar87,banavar88,curadomeunier,%
niflehilhorst92,drossel98,nogueira99,curado99,
drossel00,jorgkatzgraber1}. In spite of its
simplicity, some of the results obtained in these lattices are quite
impressive:  
(i) The bounds for the lower critical dimension, which is accepted nowadays 
to be greater than 2, but smaller than 3, were obtained on MK
lattices \cite{southernyoung} about a decade before their confirmation
through distinct numerical investigations  
\cite{bhattyoung85,ogielski85,bhattyoung88,braymoore87}. In fact, the lower
critical dimension on MK lattices was
estimated to be very close to $2.50$ \cite{curadomeunier};  
(ii) The SG critical-temperatures on the MK hierarchical
lattice of fractal dimension $D=3$, for symmetric Gaussian and bimodal 
distributions \cite{southernyoung}, 
present relative discrepancies of about $4 \% $, taking into account the 
error bars, when compared with the recent estimates from Monte Carlo
simulations on a cubic lattice~\cite{young06}. 

However, the MK lattices represent the simplest types of hierarchical
lattices and some of its results
may be quantitatively poor approximations to
well-known estimates on Bravais lattices. Other hierarchical lattices,
which present connections between such parallel paths, may represent better
approximations and in some cases, one may obtain 
precise estimates when compared to those known for Bravais lattices; this 
motivates the study of SG models on such lattices.  
In spite of this, only a few works have considered Ising SGs on hierarchical 
lattices different from those of the MK family 
\cite{nobre98,nobre00,nobre01,nobre03,nishimoriberker,salmonpre09} 
and some of them have yielded important and stimulating results:
(i) Studies on a special hierarchical lattice with fractal dimension
$D=2$ led to an estimate for the 
stiffness exponent $y$ \cite{nobre98} ($y=-1/\nu$, where
$\nu$ is the exponent associated with the divergence of the
correlation length at zero temperature) in agreement with those
obtained from other, more time-consuming, numerical 
approaches on a square lattice. An analysis of the $\pm J$ Ising SG 
model \cite{nobre01} on the same hierarchical lattice 
gave a ferromagnetic-paramagnetic critical frontier that should be a 
good approximation for the one of the corresponding model on a square lattice; 
(ii) Estimates of locations of multicritical points for SGs on some 
non-MK hierarchical lattices are in very good agreement 
with other numerical results on Bravais lattices, as well as with a
conjecture based on gauge theory \cite{nobre01,nishimoriberker}.  

In the present work we investigate Ising SGs defined on WB
hierarchical lattices through the Hamiltonian,   

\begin{equation}
\label{eq:hamiltzerofield}
{\cal H} = - \displaystyle\sum_{<ij>}J_{ij}S_{i}S_{j}
 \quad (S_{i}= \pm 1).  
\end{equation}

\vskip \baselineskip
\noindent
The lattices generated by the
cells shown in Figs.~\ref{fig:hlattice}(a), (b), and (c) present fractal
dimensions $D=2.32,3.58,$ and $4.81$; one should remind that these cells  
may be obtained from simple constructions of parts of
Bravais lattices, namely, the square, cubic and four-dimensional
hypercubic lattices, respectively~\cite{tsallismagalhaes}.
The $\{ J_{ij} \} $ denote random couplings between two spins located at
nearest-neighboring sites $i$ and $j$ of this hierarchical lattice and may
follow either the Gaussian, or bimodal ($\pm J$) distributions,     

\begin{equation} 
\label{eq:gausscoup}
P(J_{ij})= \frac{1}{\sqrt{2\pi J^{2}}}
\ \exp\left[-\frac{(J_{ij}-J_{0})^{2}}{2 J^{2}} \right]~,  
\end{equation}

\begin{equation}
\label{eq:bimcoup}
P(J_{ij})=p \ \delta(J_{ij}-J)+(1-p) \ \delta(J_{ij}+J)~. 
\end{equation}

\vskip \baselineskip
\noindent
Herein, we will restrict ourselves to those parts of the phase diagrams
associated with ferromagnetic and SG orderings only, i.e., $J_{0} \geq 0$ 
in Eq.~(\ref{eq:gausscoup}), or $1/2 \leq p \leq 1$ in
Eq.~(\ref{eq:bimcoup}). 

The RG procedure works in the inverse way of the lattice
generation, i.e., through a decimation of the internal sites of 
a given cell, leading to renormalized quantities associated with the
external sites. Defining the dimensionless couplings,
$K_{ij}=\beta J_{ij}$ [$\beta=1/(k_{B}T)$], 
the corresponding RG equations may be written in the
general form,

\begin{equation}
\label{eq:kpzerofield} 
K^{'}_{\mu \nu}=\frac{1}{4}
\log\left(\frac{Z_{--} \ Z_{++} }{Z_{-+} \ Z_{+-}}
\right)~,   
\end{equation}

\vskip \baselineskip
\noindent
where $Z_{S_{\mu},S_{\nu}}$ represent partition functions associated 
with the Hamiltonian ${\cal H}$ for a given unit cell  
with the external spins kept fixed $(S_{\mu},S_{\nu}=\pm 1)$,

\begin{equation} 
Z_{S_{\mu},S_{\nu}}= {\rm Tr}_{\{S_{i} \ (i\neq \mu,\nu)\}}
\ [\exp(-\beta {\cal H})]~.   
\end{equation}

\vskip \baselineskip
\noindent
The RG scheme is carried by following numerically
the probability distribution associated with the the dimensionless couplings 
$\{ K_{ij} \} $ \cite{southernyoung}. Operationally, this probability
distribution is 
represented by a pool of $M$ real numbers ($M$ is kept fixed throughout the
whole RG procedure), from which one may
compute its associated moments, at each renormalization step; in the 
limit  $M \rightarrow \infty$ these moments should approach those of the
distribution associated with $\{ K_{ij} \} $. The process starts by
creating a pool  
with $M$ coupling constants $\{ J_{ij} \} $ generated according to one of
the distributions of Eqs.~(\ref{eq:gausscoup}) or (\ref{eq:bimcoup}),
yielding  
an initial pool of dimensionless couplings, 
$\{ K_{ij} \} = \beta \{ J_{ij} \} $, for a given temperature. An iteration 
consists in $M$
operations, where in each of them one picks randomly a set of numbers 
from the pool (each number is assigned to a bond for a given cell 
in Fig.~\ref{fig:hlattice}) in order to
generate the effective coupling according to \eq{eq:kpzerofield}, which
will correspond to an element of the new pool. Following this procedure,  
one gets a new pool with the same size $M$ of the previous one,
representing the renormalized probability distribution.
During the RG procedure, the average, 
$<K_{ij}>$, and the width, $\sigma_{K}=<(K_{ij}-<K_{ij}>)^{2}>^{1/2}$, are 
of particular interest for the identification of the phases, in such a way
that one may obtain
the Paramagnetic ({\bf P}), Ferromagnetic ({\bf F}), and 
Spin-Glass ({\bf SG}) phases, as dominated by the attractors, 

\begin{equation}
\begin{array}{lll}
\label{eq:phasesid}
& <K_{ij}> \ \rightarrow 0~;  \ \ \ \sigma_{K}  \rightarrow 0~;  \qquad \qquad  
& {\rm {\bf P} \ phase}~, \cr
& <K_{ij}> \ \rightarrow \infty~;  \ \ \sigma_{K}  \rightarrow \infty \ \
(<K_{ij}>/ \sigma_{K}  \rightarrow \infty)~;   \qquad \qquad  
& {\rm {\bf F} \ phase}~, \cr
& <K_{ij}> \ \rightarrow 0~;   \ \ \ \sigma_{K}  \rightarrow \infty~;   \qquad \qquad  
& {\rm {\bf SG} \ phase}~. 
 \end{array}
 \end{equation} 

\vskip \baselineskip
\noindent
Strictly speaking, this procedure should be carried for many different
initial pools of real numbers, over which one may compute sample 
averages. However, one 
may also get good critical-frontier estimates by analyzing a 
sufficiently large single pool; 
the results that follow were obtained by considering a single pool of size
$M=400000$ real numbers.   

\begin{figure}[t]
\begin{center}
\includegraphics[width=0.60\textwidth,angle=0]{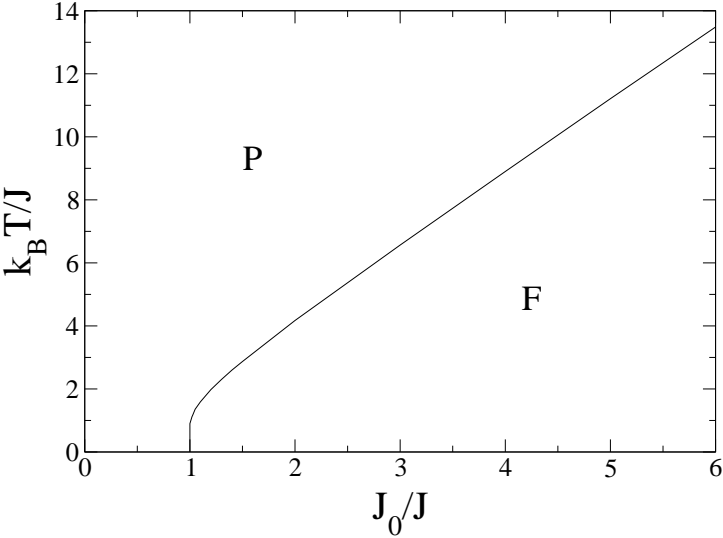}
\end{center}
\protect\caption{Phase diagram for an Ising SG on the hierarchical 
lattice defined by the unit cell of Fig.~\ref{fig:hlattice}(a), with a
Gaussian distribution for the couplings.} 
\label{fig:phasediaggaussh02d}
\end{figure}

As expected, we did not find a finite-temperature SG phase for the
hierarchical lattice defined by the unit cell of
Fig.~\ref{fig:hlattice}(a), for either one of the  distributions of
Eqs.~(\ref{eq:gausscoup}) or (\ref{eq:bimcoup}).  
The phase diagram for the case of the  
Gaussian distribution for the couplings is exhibited in 
Fig.~\ref{fig:phasediaggaussh02d}, where only two phases are present, 
namely, the {\bf P} and {\bf F} phases. 
The points used in order to draw this phase diagram were
calculated from the standard narrowing RG procedure, with at least
a two-decimal-digit certainty (error bars on third decimals). 
Due to the duality property of this 
unit cell \cite{tsallismagalhaes}, one should obtain the exact critical
temperature of the ferromagnetic Ising model on the square lattice from the
phase diagram of  
Fig.~\ref{fig:phasediaggaussh02d}; indeed, computing the 
slope of the critical frontier of Fig.~\ref{fig:phasediaggaussh02d} 
for $(J_{0}/J)$ large, one gets $(k_{\rm B}T_{c}/J_{0})=2.270 \pm 0.002$.
Although this model does not exhibit a SG phase at finite temperatures, one
may still compute the stiffness exponent $y$, which rules the
zero-temperature scaling
behavior of the width of a continuous coupling distribution associated with
blocks of linear size $L$~\cite{braymoore87}, 

\begin{equation} 
J^{\prime} (L) \sim J L^{y}~; \quad L=2^{\cal N}~; 
\quad J=<(J_{ij}-<J_{ij}>)^{2}>^{1/2}~.  
\end{equation} 

\vskip \baselineskip
\noindent 
For sufficiently small values of $(J_{0}/J)$, the sign of the stiffness
exponent $y$ is directly associated with the low-temperature phase; for a
positive (negative) $y$ the system scales to strong (weak) couplings,
characteristic of a SG (paramagnetic) state at low temperatures.
Therefore, one expects $y<0$ for the WB hierachical lattice of 
Fig.~\ref{fig:hlattice}(a), whereas one should get $y>0$ for those of  
Figs.~\ref{fig:hlattice}(b) and~\ref{fig:hlattice}(c). In the first case
one may use the scaling relation, $\nu = - 1/y$~\cite{braymoore87}, for  
a phase transition
in the limit $T \rightarrow 0$, where $\nu$ is the exponent associated with
the divergence of the correlation length, 
$\xi \sim T^{-\nu}$. We have computed $y=-0.290 \pm 0.003$, leading to 
$\nu=3.45 \pm 0.04$, for all $J_{0}$ in the interval 
$0 \leq (J_{0}/J)<1$ (cf. Fig.~\ref{fig:phasediaggaussh02d}). 
This estimate coincides, within the error bars, with those obtained 
for the particular case $J_{0}=0$ in Ref.~\cite{nobre98} 
for the same hierarchical lattice, as well as with several estimates for
the square lattice, like $y=-0.291 \pm 0.002$~\cite{braymoore87}, 
$y=-0.287 \pm 0.004$~\cite{hartmann}, and 
$y=-0.284 \pm 0.004$~\cite{weigel}. 
Although the lattice considered herein presents a fractal dimension 
$D \approx 2.32$, which
is about $15\% $ higher than the dimension of the square lattice, it has
been much used in the literature as an approach for models on the later,
essentially because it may be obtained through a simple construction 
from a piece of the square lattice, keeping one of its most important
properties, the self-duality~\cite{tsallismagalhaes}.
 
The phase diagram for the case of the  
hierarchical lattice defined by the unit cell of
Fig.~\ref{fig:hlattice}(a), with a 
bimodal distribution for the couplings, is qualitatively
similar to the one already investigated in Ref.~\cite{nobre01}. 
From the present 
analysis one concludes that the SG lower critical dimension in the WB
hierarchical-lattice family  should be greater than $2.32$; this result is
in agreement with calculations carried on  
MK lattices, where this lower critical dimension was estimated
to be very close to $2.50$ \cite{curadomeunier}.   

\begin{figure}[t]
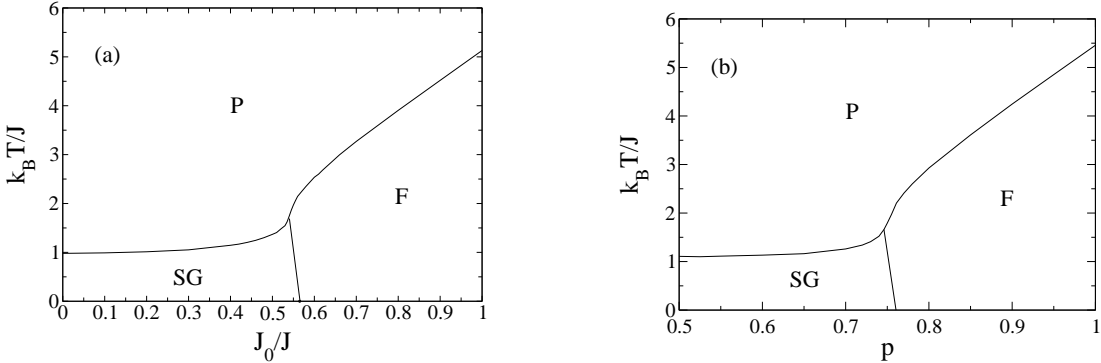

\begin{center}
\includegraphics[width=0.40\textwidth,angle=0]{fig4a.eps}
\hspace{1.5cm}
\includegraphics[width=0.40\textwidth,angle=0]{fig4b.eps}
\end{center}
\protect\caption{Phase diagrams of the Ising SG on the hierarchical 
lattice defined by the unit cell of Fig.~\ref{fig:hlattice}(b): 
(a) Gaussian  distribution for the couplings; (b) bimodal distribution for 
the couplings.}
\label{fig:phasediagsh03d}
\end{figure}

The phase diagrams for the hierarchical lattice defined by the unit cell of 
Fig.~\ref{fig:hlattice}(b) are shown in Fig.~\ref{fig:phasediagsh03d}, for
couplings following the Gaussian distribution  
[Fig.~\ref{fig:phasediagsh03d}(a)], or the bimodal distribution
[Fig.~\ref{fig:phasediagsh03d}(b)]. The coordinates of the most important
critical points of these phase diagrams are given explicitly in Tables~I
and II. 
Similarly to the previous case, the points used in order to draw these 
phase diagrams were calculated from the standard narrowing RG procedure,
with at least 
a two-decimal-digit certainty (error bars on third decimals). 
It is important to notice the estimates of the critical temperatures for
symmetric distributions, namely, $(k_{B}T_{c}/J)=0.980(2)$ and  
$(k_{B}T_{c}/J)=1.112(2)$;  
these estimates are in very good agreement with recent results from 
Monte Carlo simulations on a cubic lattice, which yielded 
$(k_{B}T_{c} /J) = 0.951(9)$ and $(k_{B}T_{c} /J) = 1.120(4)$, for
symmetric Gaussian and bimodal distributions, respectively~\cite{young06}.
Comparing the present results with those of Ref.~\cite{young06}, taking
into account the error bars, one finds a relative discrepancy of about 
$2\%$ in the Gaussian case, whereas for the symmetric bimodal distribution
the two estimates essentially coincide (leading to a relative discrepancy
of about  $0.2 \%$).
Such good agreements on critical-temperature estimates, between those of a
hierarchical lattice with fractal dimension $D \approx 3.58$ and those of 
the cubic lattice, are justified by the fact that the cell of 
Fig.~\ref{fig:hlattice}(b) may be obtained from a simple construction of a
piece of the cubic lattice~\cite{tsallismagalhaes}. 
On the other hand, the critical temperature associated with the corresponding
ferromagnetic Ising model, which may be obtained either from 
the $p=1$ critical
point separating phases {\bf P} and {\bf F} in 
Fig.~\ref{fig:phasediagsh03d}(b), or by 
computing the slope of the critical frontier of 
Fig.~\ref{fig:phasediagsh03d}(b) for $(J_{0}/J)$ large, is given by
$(k_{\rm B}T_{c}/J_{0})=5.46 \pm 0.01$, which is somewhere in between the
estimates of critical temperatures for the ferromagnetic Ising model on the
cubic [$(k_{\rm B}T_{c}/J_{0}) \approx 4.512$] and the 
four-dimensional hypercubic lattice 
[$(k_{\rm B}T_{c}/J_{0}) \approx 6.680]$, obtained
from Monte Carlo simulations~\cite{staufferbjp}. 
 
\begin{figure}[t]
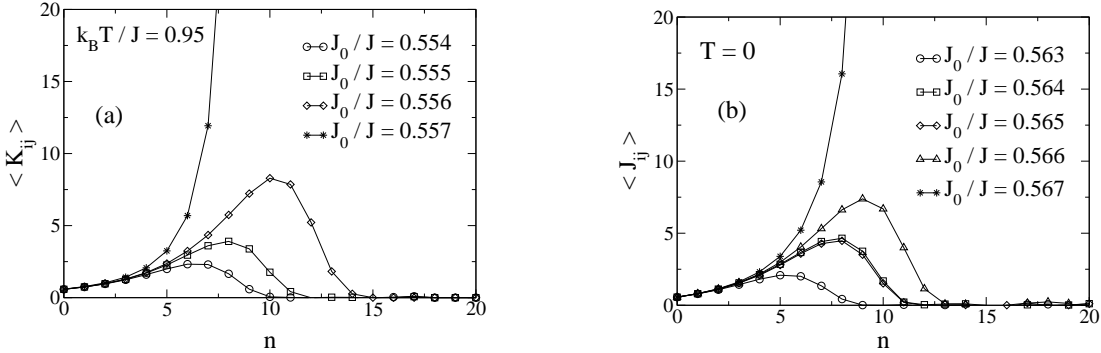

\begin{center}
\includegraphics[width=0.40\textwidth,angle=0]{fig5a.eps}
\hspace{1.5cm}
\includegraphics[width=0.40\textwidth,angle=0]{fig5b.eps}
\end{center}
\protect\caption{Evolution of the average of couplings with 
the RG step $n$, 
for the case of an initial Gaussian distribution, 
considering typical values of $J_{0}/J$ around the critical frontier 
separating the phases {\bf SG} and {\bf F}. (a) Fixed temperature below
the SG critical temperature; (b) Zero temperature.}  
\label{fig:j0evolutionh03d}
\end{figure}

Another important aspect of the phase diagrams shown in 
Fig.~\ref{fig:phasediagsh03d} is a small reentrance effect in the region
slightly to the right of the multicritical point; by lowering the
temperature, one may go from a high-temperature disordered ({\bf P})
phase to an ordered one ({\bf F} phase) and then, back to a more
disordered ({\bf SG}) phase at low temperatures. By comparing the
coordinates of the multicritical points and the respective zero-temperature 
ones in Tables~I and II, one sees that this effect, although very weak, is 
outside the error bars of the method.  
This is illustrated in Fig.~\ref{fig:j0evolutionh03d}, which shows the
evolution of the average $<K_{ij}>$ with the RG step $n$, in the case of 
couplings following an initial Gaussian distribution, 
for typical values of $J_{0}/J$ around the critical frontier 
separating the phases {\bf SG} and {\bf F}. In  
Fig.~\ref{fig:j0evolutionh03d}(a) one sees the behavior of $<K_{ij}>$ for a
fixed temperature [$(k_{B}T/J)=0.95$], below the SG critical temperature,
showing that for $(J_{0}/J)=0.556$ one is still on the {\bf SG} phase, 
whereas $<K_{ij}>$ diverges for $(J_{0}/J)=0.557$, signalling a
{\bf F} phase. The same procedure is applied to zero temperature in 
Fig.~\ref{fig:j0evolutionh03d}(b), where one sees these two different types
of behavior for $(J_{0}/J)=0.566$ and $(J_{0}/J)=0.567$, respectively. 
It should be mentioned that this type of reentrance phenomena has been 
observed experimentally, e.g., in ${\rm Eu_{x}Sr_{1-x}S}$ \cite{malettaconvert}
and ${\rm AuFe}$ ($14 \% $ Fe) \cite{craneclaus}. Theoretically, it has 
been also found in the mean-field replica-symmetric solution (which turns
out to be unstable at low temperatures) of the Ising SG, 
although its correct solution, 
characterized by replica-symmetry breaking, has washed away the 
reentrance~\cite{dotsenkobook,nishimoribook,parisirsb}. 
It is important to stress that reentrance phenomena on theoretical SG
models are very subtle and
difficult to be obtained numerically, in such a way that 
only a few works in the literature were able to capture such an effect
so far (see, e.g., Refs.~\cite{nobre01,toldin}). 

\begin{table}
\begin{center}
\begin{tabular}{||c|c|c|c||}			 \hline

Hierarchical & $k_{B}T_{c}/J$ 
& $J_{0}/J$ & Multicritical Point \\
lattice & ($J_{0}=0$) & ($T=0$) &  \\     \hline

Cell 1(b) & $0.980(2)$ & $0.5665(5)$ & $(k_{B}T_{c}/J)=1.690(2)$; 
$(J_{0}/J)=0.538(2)$ \\   

Cell 1(c) & $2.35(1)$ & $0.374(3)$ & $(k_{B}T_{c}/J)=2.77(1)$; 
$(J_{0}/J)=0.35(1)$ \\   \hline
\end{tabular}
\end{center}
\caption{\small
Values of typical critical points in the phase diagrams of the 
Ising SG, on
the hierarchical lattices defined by unit cells of
Figs.~\ref{fig:hlattice}(b) and \ref{fig:hlattice}(c), for couplings
following a Gaussian distribution [Eq.~(\ref{eq:gausscoup})]: critical
temperature for a symmetric distribution, zero-temperature critical point
separating phases {\bf SG} and {\bf F}, and coordinates of the multicritical
point where the three critical frontiers meet. The
error bars refer to the usual approach to criticality characteristic of the
RG technique.} 
\end{table}

We have also computed the zero-temperature stiffness exponent 
$y$ for the phase diagram of 
Fig.~\ref{fig:phasediagsh03d}(a). As expected, a positive value,
$y=0.297 \pm 0.003$, signals a SG phase at finite temperatures;
furthermore, this estimate is universal for all values of $J_{0}$ within 
the SG phase. This value is close to the recent one obtained for a 
MK lattice of fractal dimension $D=3$ 
($y \approx 0.27$~\cite{jorgkatzgraber1}), but
it is slightly larger than those for the cubic lattice, 
$y=0.19 \pm 0.01$~\cite{braymoore87} and 
$y=0.20 \pm 0.05$~\cite{jorgkatzgraber1}.

The phase diagrams for the hierarchical lattice defined by the unit cell of 
Fig.~\ref{fig:hlattice}(c), for couplings following either initial
Gaussian of bimodal distributions, are qualitatively similar to those shown
in Figs.~\ref{fig:phasediagsh03d}(a) and~\ref{fig:phasediagsh03d}(b),
respectively. Values of typical critical points in these phase 
diagrams are given explicitly in Tables~I and II.
The higher complexity of this lattice generates larger errors in the
numerical estimates; however, the points in such phase diagrams 
were computed with at least
a one-decimal-digit certainty (error bars on second decimals). 
In particular, one should mention the estimates of the critical
temperatures for 
symmetric distributions, namely, $(k_{B}T_{c}/J)=2.35(1)$ and  
$(k_{B}T_{c}/J)=2.515(2)$, which are about $25 \%$ above those
obtained from Monte Carlo simulations for the four-dimensional hypercubic
lattice, i.e., $(k_{B}T_{c}/J) \approx 1.80$, for the Gaussian
distribution~\cite{jorgkatzgraber2}, and $(k_{B}T_{c}/J) \approx 2.00$, 
for the bimodal distribution~\cite{hukushima}, respectively. 
These overestimates are
consistent with the fact that the fractal dimension of the lattice
considered herein is well above four ($D \approx 4.81$). Although we are
not aware of numerical simulations carried for Ising SGs on lattices of
higher dimensions, the present 
estimate for the symmetric bimodal distribution is slightly below 
the one obtained through series expansions on the 
five-dimensional hypercubic lattice, 
$(k_{B}T_{c}/J) \approx 2.57$~\cite{klein}. 

\begin{table}
\begin{center}
\begin{tabular}{||c|c|c|c||}			 \hline

Hierarchical & $k_{B}T_{c}/J$ 
&  $p_{c}$ & Multicritical Point \\
lattice & ($p=1/2$) & ($T=0$) & \\     \hline

Cell 1(b) & $1.112(2)$ & $0.760(1)$ & $(k_{B}T_{c}/J)=1.620(2)$; 
$p_{c}=0.745(2)$ \\   

Cell 1(c) & $2.515(2)$ & $0.667(2)$ & $(k_{B}T_{c}/J)=2.836(2)$; 
$p_{c}=0.664(2)$ \\   \hline
\end{tabular}
\end{center}
\caption{\small
Values of typical critical points in the phase diagrams of the Ising SG, on
the hierarchical lattices defined by unit cells of 
Figs.~\ref{fig:hlattice}(b) and \ref{fig:hlattice}(c), for couplings
following a bimodal distribution [Eq.~(\ref{eq:bimcoup})]: critical
temperature for a symmetric distribution, zero-temperature critical point
separating phases {\bf SG} and {\bf F}, and coordinates of the multicritical
point where the three critical frontiers meet. The
error bars refer to the usual approach to criticality characteristic of the
RG technique.} 
\end{table}

To conclude, we have investigated nearest-neighbor-interaction Ising 
spin glasses defined on three different hierarchical lattices, 
belonging to a family of
Wheatstone-Bridge lattices, characterized by fractal
dimensions $D \approx 2.32,3.58,$ and $4.81$. 
The interactions among pairs of
spins were chosen from either a Gaussian, or a bimodal ($\pm J$),
non-centered distribution.    
Through calculations of the stiffness exponent, we have shown that 
the spin-glass lower
critical dimension in these lattices should be greater than $2.32$~. 
Finite-temperature spin-glass phases were found for   
the lattices of fractal dimension $D \approx 3.58$ and $D \approx 4.81$. In
the former case, whose hierarchical lattice is constructed as an
approximation to the cubic lattice,
the estimates of spin-glass critical temperatures 
associated with zero-centered  
Gaussian and bimodal distributions are very close to the 
most recent estimates from
extensive numerical simulations carried on the cubic lattice. 
Phase diagrams were obtained for
couplings following non-centered distributions -- which are 
not entirely known, up to the moment, for spin-glass models on
Bravais lattices -- either for 
$J_{0} \geq 0$ in the case of
the Gaussian distribution, or $1/2 \leq p \leq 1$ in the case of the
bimodal distribution; these phase diagrams are expected to represent good
approximations for the corresponding models on Bravais lattices. 

\vskip 2\baselineskip

{\large\bf Acknowledgments}

\vskip \baselineskip
\noindent
We thank Prof. E.~M.~F. Curado for fruitful
conversations. The partial financial supports from
CNPq and Pronex/MCT/FAPERJ (Brazilian agencies) are acknowledged. 

\vskip 2\baselineskip

\newpage

\begin{center}

{\large\bf Figure Captions}

\end{center}

\noindent
{\bf Fig. 1:} Basic cells of the WB family of
hierarchical lattices, with a scaling factor $b=2$.  
In each case, the empty circles ($\mu$ and $\nu$) represent the external
sites  
of the cell, whereas the black circles are internal sites to be 
decimated in the RG procedure. 
(a) The WB cell of fractal dimension $D=(\ln5)/(\ln2) \approx 2.32$; 
(b) The WB cell of fractal dimension $D=(\ln12)/(\ln2) \approx 3.58$; 
(c) The WB cell of fractal dimension $D=(\ln28)/(\ln2) \approx 4.81$. For
clearness, all the bonds of cell (c) are not shown explicitly; the external  
site $\mu$ is connected directly to sites $5,6,7,$ and $8$ (as illustrated
by the bond connecting sites $\mu$ and $5$, whereas site $\nu$ is connected
directly to sites $1,2,3,$ and $4$ (as illustrated by the bond connecting
sites $\nu$ and $3$).

\vskip \baselineskip
\noindent
{\bf Fig. 2:} First three levels (levels $0,1,$ and $2$) in the
generation process of the Wheatstone-Bridge hierarchical lattice with
fractal dimension $D \approx 2.32$. The lattice-generation process starts at 
the $0$-th level with a single bond and at each step a bond is replaced by
its respective basic cell. In each level, the empty circles ($\mu$ and $\nu$) 
represent the external sites of the lattice, 
whereas the black circles are internal sites to be decimated in successive 
RG operations.

\vskip \baselineskip
\noindent
{\bf Fig. 3:} Phase diagram for an Ising SG on the hierarchical 
lattice defined by the unit cell of Fig.~\ref{fig:hlattice}(a), with a
Gaussian distribution for the couplings. 
 
\vskip \baselineskip
\noindent
{\bf Fig. 4:} Phase diagrams of the Ising SG on the hierarchical 
lattice defined by the unit cell of Fig.~\ref{fig:hlattice}(b): 
(a) Gaussian  distribution for the couplings; (b) bimodal distribution for 
the couplings.

\vskip \baselineskip
\noindent
{\bf Fig. 5:} Evolution of the average of couplings with 
the RG step $n$, 
for the case of an initial Gaussian distribution, 
considering typical values of $J_{0}/J$ around the critical frontier 
separating the phases {\bf SG} and {\bf F}. (a) Fixed temperature below
the SG critical temperature; (b) Zero temperature.

\newpage

\begin{thebibliography}{40}

\bibitem{dotsenkobook}
V. Dotsenko, Introduction to the Replica Theory of Disordered
Statistical Systems, Cambridge University Press, Cambridge, 2001.

\bibitem{nishimoribook}
H. Nishimori, Statistical Physics of Spin Glasses and Information
Processing, Oxford University Press, Oxford 2001.

\bibitem{parisirsb}
G. Parisi, Phys. Rev. Lett. {\bf 43}~, 1754-1756 (1979); {\bf 50} (1983) 1946-1948.

\bibitem{parisireview}
E. Marinari, G. Parisi, and J.J. Ruiz-Lorenzo, in: A.~P. Young (Ed.), 
Spin Glasses and Random Fields, World Scientific, Singapore, 1998, 
pp. 59-98.

\bibitem{southernyoung}
B.~W. Southern and A.~P. Young, J. Phys. C {\bf 10} (1977), 2179-2195.

\bibitem{bhattyoung85}
R.~N. Bhatt and A.~P. Young, Phys. Rev. Lett. {\bf 54}~(1985), 924-927. 

\bibitem{ogielski85}
A.~T. Ogielski and I. Morgenstern, Phys. Rev. Lett. {\bf 54}~(1985) 
928-931. 

\bibitem{braymoore87}
A.~J. Bray and M.~A. Moore, in: J.~L. van Hemmen and I. Morgenstern (Eds.),
Heidelberg Colloquium on Glassy Dynamics, Lecture Notes in Physics {\bf 275}, 
Springer-Verlag, Berlin, 1987, 121-153. 

\bibitem{bhattyoung88}
R.~N. Bhatt and A.~P. Young, Phys. Rev. B {\bf 37} (1988) 5606-5614.

\bibitem{young04}
A.~P. Young and H.~G. Katzgraber, Phys. Rev. Lett. {\bf 93} (2004) 207203-1-207203-4.

\bibitem{young06}
H.~G. Katzgraber, M. K\"orner, and A.~P. Young, Phys. Rev. B {\bf 73} (2006) 
224432-1- 224432-11.

\bibitem{jorg}
T. J\"org, H.~G. Katzgraber, and F. Krz\c{a}kala,
Phys. Rev. Lett. {\bf 100} (2008) 197202-1-197202-4.

\bibitem{jorgkatzgraber1}
T. J\"org and H.~G. Katzgraber, Phys. Rev. Lett. {\bf 101} (2008)
197205-1-197205-4. 

\bibitem{tsallismagalhaes}
C. Tsallis and A.~C.~N. de Magalh\~aes, Phys. Rep. {\bf 268} (1996) 305-430. 

\bibitem{banavar87}
J.~R. Banavar and A.~J. Bray, Phys. Rev. B {\bf 35}, 8888-8890 (1987).

\bibitem{banavar88}
J.~R. Banavar and A.~J. Bray, Phys. Rev. B {\bf 38} (1988) 2564-2569.

\bibitem{curadomeunier}
E.~M.~F. Curado and J.~L. Meunier, Physica A {\bf 149} (1988) 164-181.

\bibitem{niflehilhorst92}
M. Nifle and H.~J. Hilhorst, Phys. Rev. Lett. {\bf 68} (1992) 2992-2995.

\bibitem{nobre98}
F.~D. Nobre, Phys. Lett. A {\bf 250} (1998) 163-169.

\bibitem{drossel98}
M.~A. Moore, H. Bokil, and B. Drossel, Phys. Rev. Lett. {\bf 81} (1998) 4252-4255.

\bibitem{nogueira99}
E. Nogueira Jr., S. Coutinho, F.~D. Nobre, and E.~M.~F. Curado, Physica A
{\bf 271} (1999) 125-132.

\bibitem{curado99}
E.~M.~F. Curado, F.~D. Nobre, and S. Coutinho, Phys. Rev. E {\bf 60} (1999)
3761-3770.

\bibitem{nobre00}
F.~D. Nobre, Physica A {\bf 280} (2000) 456-464.

\bibitem{drossel00}
B. Drossel, H. Bokil, M.~A. Moore, and A.~J. Bray, 
Eur. Phys. J. B {\bf 13} (2000) 369-375. 

\bibitem{nobre01}
F.~D. Nobre, Phys. Rev. E {\bf 64} (2001) 046108-1-046108-6.

\bibitem{nobre03}
F.~D. Nobre, Physica A {\bf 319} (2003) 362-370.

\bibitem{nishimoriberker}
M. Ohzeki, H. Nishimori, and A.~N. Berker, 
Phys. Rev. E {\bf 77} (2008), 061116-1-061116-11.

\bibitem{salmonpre09}
O.~R. Salmon and F.~D. Nobre, Phys. Rev. E {\bf 79} (2009) 051122-1-051122-6.

\bibitem{migdal}
A.~A. Migdal, Sov. Phys. JETP {\bf 42} (1976) 743-746.

\bibitem{kadanoff}
L.~P. Kadanoff, Ann. Phys. {\bf 100} (1976) 359-394. 

\bibitem{hartmann}
A.~K. Hartmann, A.~J. Bray, A.~C. Carter, M.~A. Moore, and A.~P. Young, 
Phys. Rev. B {\bf 66} (2002) 224401-1-224401-4.

\bibitem{weigel}
M. Weigel and D. Johnston, Phys. Rev. B {\bf 76} (2007) 054408-1-054408-8.

\bibitem{staufferbjp}
D. Stauffer, Braz. J. Phys. {\bf 30} (2000) 787-793. 

\bibitem{malettaconvert}
H. Maletta and P. Convert, Phys. Rev. Lett. {\bf 42} (1979) 108-111. 

\bibitem{craneclaus}
S. Crane and H. Claus, Phys. Rev. Lett. {\bf 46} (1981) 1693-1695.

\bibitem{toldin}
F.~P. Toldin, A. Pelissetto, and E. Vicari, J. Stat. Phys. {\bf 135} (2009)
1039-1061. 

\bibitem{jorgkatzgraber2}
T. J\"org and H.~G. Katzgraber, Phys. Rev. B {\bf 77} (2008) 214426-1-214426-12.

\bibitem{hukushima}
K. Hukushima, Phys. Rev. E {\bf 60} (1999) 3606-3613.

\bibitem{klein}
L. Klein, J. Adler, A. Aharony, A.~B. Harris, and Y. Meir, 
Phys. Rev. B {\bf 43} (1991) 11249-11273. 

\end{thebibliography}
\end{document}